\documentclass[%
 reprint,
 amsmath,amssymb,
 aps,
]{revtex4-2}
\usepackage{graphicx}
\usepackage{dcolumn}
\usepackage{bm}
\usepackage{braket}
\usepackage{array}
\usepackage{multirow}
\usepackage{url}
\usepackage{tabularx}
\usepackage[bookmarks=true,colorlinks=true,urlcolor=blue,linkcolor=blue,citecolor=blue]{hyperref}
\usepackage{longtable}
\usepackage{makecell}
\usepackage{siunitx}
\usepackage{physics}
\usepackage{float}
\usepackage{boldline}
\usepackage{tabularx}
\usepackage{adjustbox}
\usepackage{cancel}
\usepackage[dvipsnames]{xcolor}
\usepackage{ulem}
\begin{document}

\preprint{APS/123-QED}

\title{Quantum study of ultracold atom-ion excitation exchange}

\author{Tibor Jónás}
\affiliation{University of Debrecen, Doctoral School of Physics, Egyetem tér 1., 4032 Debrecen, Hungary}
\affiliation{Université Paris-Saclay, CNRS, Laboratoire Aimé Cotton, Orsay, 91400, France}
\affiliation{HUN-REN Institute for Nuclear Research (HUN-REN ATOMKI), Bem tér 18/c, 4026 Debrecen, Hungary}
\author{Romain Vexiau}
\affiliation{Université Paris-Saclay, CNRS, Laboratoire Aimé Cotton, Orsay, 91400, France}
\author{Nadia Bouloufa-Maafa}
\affiliation{Université Paris-Saclay, CNRS, Laboratoire Aimé Cotton, Orsay, 91400, France}
\author{Eliane Luc-Koenig}
\affiliation{Université Paris-Saclay, CNRS, Laboratoire Aimé Cotton, Orsay, 91400, France}
\author{Andrea Orbán}
\affiliation{HUN-REN Institute for Nuclear Research (HUN-REN ATOMKI), Bem tér 18/c, 4026 Debrecen, Hungary}
\author{Olivier Dulieu}
\affiliation{Université Paris-Saclay, CNRS, Laboratoire Aimé Cotton, Orsay, 91400, France}

\date{\today}


\begin{abstract} 
The quantum dynamics of ultracold collisions between rubidium atoms and excited metastable strontium ions is treated in the laboratory frame, enlightening the importance of the coupling between internal angular momenta of the particles and their mutual rotation. The study reveals a subtle competition between electronic excitation exchange and fine structure quenching, with no charge exchange, which is found to be very sensitive to the details of ion-atom interactions. The rate constant for electronic excitation exchange is found in agreement with the experimental results of Ben-Shlomi \textit{et al.} (Phys. Rev. A \textbf{102}, 031301(R) (2020)), while the rate for fine structure quenching is predicted to strongly depend on the initial polarization of the reactants. 
\end{abstract}

\pacs{....}
\maketitle

\section{Introduction}
Research on laser cooling and trapping of quantum particles down to ultralow kinetic energies (equivalent to temperatures well below 1 millikelvin) progressed in parallel for neutral atoms \cite{phillips1998,cohen-tannoudji1998b,chu1998} and atomic ions \cite{Wineland2013}, and led to ground-breaking achievements like the realization of quantum degenerate gases \cite{cornell2002,ketterle2002} and single-ion manipulation \cite{Wineland2013}. About 25 years ago emerged the idea of combining these two experimental approaches to study the quantum nature of ion-atom interactions in devices now known as hybrid traps \cite{makarov2003}. Indeed, unlike ultracold neutral atomic gases where the dynamics is governed by the van der Waals interaction (varying as $R^{-6}$) between atoms at large distances $R$, the ion-atom interaction is dominated by a charge-induced dipole interaction varying as $R^{-4}$. Reaching the quantum regime (\textit{i.e.} scattering regime in which only a few partial waves, and ultimately a single partial wave, contribute) in such a case remains an ongoing challenge as temperatures much lower than for neutral atom gases are needed. Such hybrid systems have important potential applications \cite{tomza2019}, for example, for studying charge transport or impurities in a quantum gas, thus providing a promising platform for quantum simulation. Such studies can also be extended to ultracold chemistry, involving, for instance, an atomic ion and a polar molecule, giving rise to even stronger long-range interactions (charge-dipole interaction varying as $R^{-2}$), which might compete with the dipole-dipole interaction (varying as $R^{-3}$) between polar molecules \cite{karpa2025}.

Among the various combinations of atoms and ions that have been experimentally addressed \cite{tomza2019}, bringing alkali-metal (AM) atoms and alkaline-earth (AE) ions inside a hybrid trap has been found to be very convenient, as both species can be routinely laser-cooled. Moreover, resonant charge exchange cannot occur, so that other inelastic processes involving excitation transfer or non-resonant charge exchange can be accurately observed. Despite the fact that the sum of ground state energies of a given AM-AE$^+$ pair is larger than that of the AM$^+$-AE (except for the notable example of the Li-Ba$^+$ pair), radiative charge exchange is predicted with very low probability \cite{dasilva2015}. Another common feature of interacting pairs AM-AE$^+$ is that the laser-cooling scheme of AE$^+$ inherently involves its lowest $^2D$ metastable excited energy level. Therefore, internal energy can be selectively provided to the system, thus allowing for enhanced reactivity of the pair. The excited electronic states of the corresponding ions (AM-AE)$^+$ contribute to the dynamics, strongly constraining the associated theoretical models. Several combinations have been experimentally investigated along this line, including Rb-Ca$^+$ \cite{hall2011,hall2013a}, Rb-Sr$^+$ \cite{benshlomi2020}, Rb-Ba$^+$ \cite{hall2013b,mohammadi2021}, Rb-Yb$^+$ \cite{ratschbacher2012}, Li-Ca$^+$ \cite{haze2015,saito2017}, Li-Ba$^+$ \cite{xing2024}, Li-Yb$^+$ \cite{joger2017}. 

\begin{figure}[h!]
 \centering
 \includegraphics[width=1.0\columnwidth]{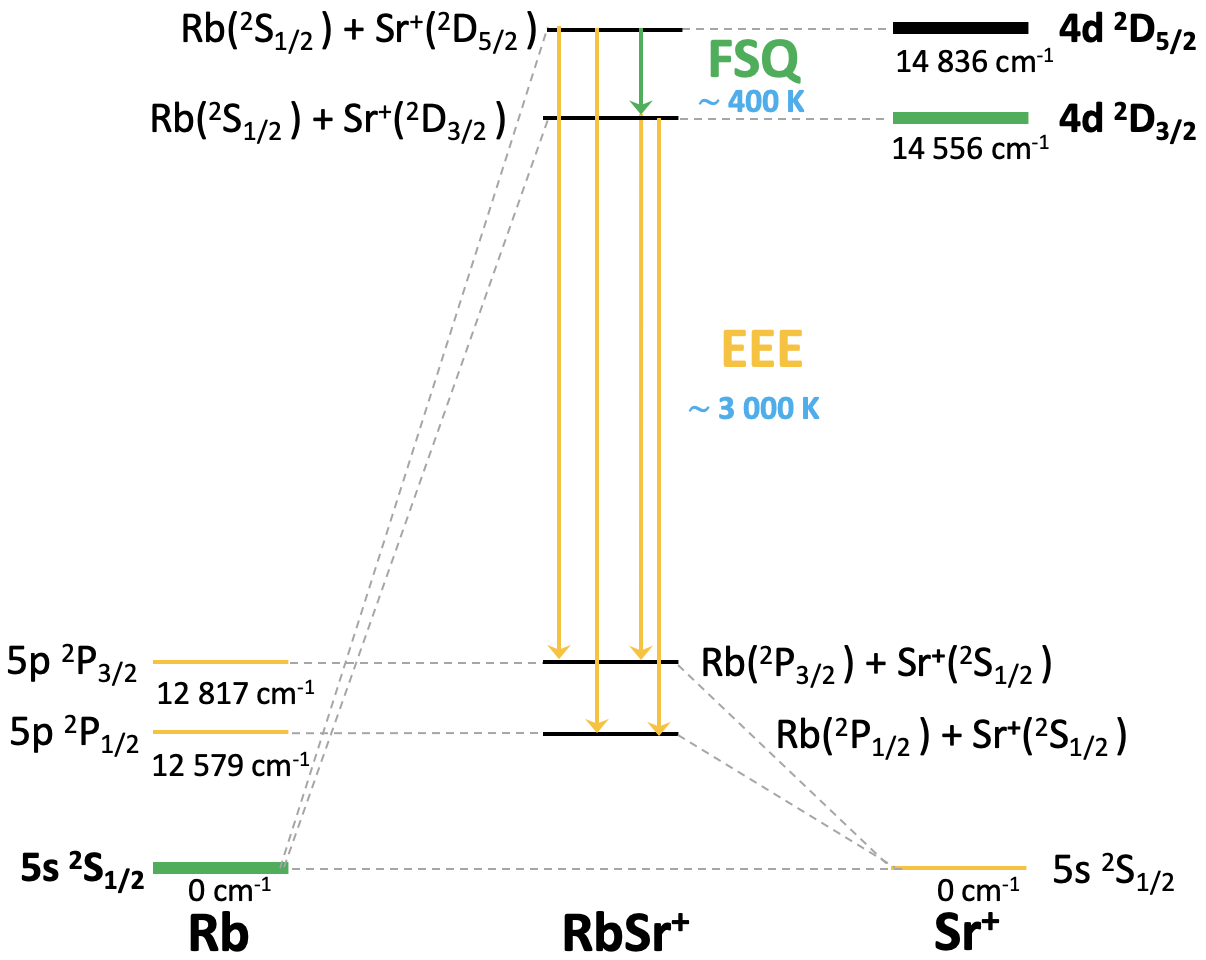}
 \caption{Energy levels of Rb and Sr$^+$ relevant for the present study, with their excitation energy (not to scale). The associated incoming and outgoing channels for Eqs. \ref{eq:EEE}, \ref{eq:FSQ} are drawn with orange and green arrows indicating the EEE and FSQ processes, respectively. The value of released kinetic energies for each process measured in the experiment (see paper I)  and expressed as a temperature is displayed. }
 \label{fig:ionatomlevels}
 \end{figure}

In a joint experimental and theoretical study of ultracold collisions between ground state $^{87}$Rb($5s\,^2S_{1/2}$) atoms and excited metastable $^{88}$Sr$^+$($4d\,^2D_{3/2,5/2}$) ions (\cite{benshlomi2020} hereafter referred to as paper I), it has been shown that two processes were competing with each other: electronic excitation exchange (EEE)
\begin{eqnarray}
\text{Rb}(5s \; ^2S_{1/2}) &+& \text{Sr}^+(4d \; ^2D_{3/2,5/2})  \to \nonumber \\
&\to&\text{Rb}(5p \; ^2P_{1/2,3/2}) + \text{Sr}^+(5s \; ^2S_{1/2}),
\label{eq:EEE}
\end{eqnarray}
and fine-structure quenching (FSQ)
\begin{eqnarray}
\text{Rb}(5s \; ^2S_{1/2}) &+& \text{Sr}^+(4d \; ^2D_{5/2})  \to \nonumber \\
&\to&\text{Rb}(5s \; ^2S_{1/2}) + \text{Sr}^+(4d \; ^2D_{3/2}),
\label{eq:FSQ}
\end{eqnarray}
without contribution from radiative (RCE) or non-radiative charge exchange (NRCE). In contrast, RCE was found to be active \cite{sikorsky2018} but rare \cite{dasilva2015} when both particles are in their ground state. Figure  \ref{fig:ionatomlevels} displays the relevant energy levels and energies released by the processes. The theoretical model of paper I revealed the complexity of the potential energy curves (PECs) of the RbSr$^+$ ion, and the role of the strong Coriolis-spin mixing in the observed non-radiative processes, which were qualitatively interpreted with a semiclassical Landau-Zener (LZ) model. However, a quantitative agreement with the experimental reaction rates was not obtained.

In the present paper, following our earlier study on Li-Ba$^+$ \cite{xing2024}, we develop a full quantum scattering description for the collision Rb($5s\,^2S_{1/2}$)+Sr$^+$($4d\,^2D_{3/2,5/2}$), based on the PECs determined in paper I. Special care is brought to the description of spin-orbit coupling (SOC), which competes with other non-adiabatic effects, requiring a specific quasi-diabatic approach with two steps.

The paper is organized as follows: Section \ref{sec:PECSBF} recalls the main steps of our calculations of the Hund's case (a) PECs, including their extrapolation at large distances, and the adopted two-state model for non-adiabatic couplings. In Section \ref{sec:soc-diab}, the implementation of the molecular spin-orbit interaction is described, relying on a two-step quasi-diabatic approach involving atomic spin-orbit interaction. The dynamics of the cold collision between ground state Rb atoms and metastable Sr$^+$ ions is presented in Section \ref{sec:dynamic}, and the resulting computed branching ratios of these collisions are discussed in Section \ref{sec:branching}. Concluding remarks and perspectives are proposed in Section \ref{sec:conclusion}.

\section{Molecular structure of R\lowercase{b}S\lowercase{r}$^+$}
\label{sec:PECSBF}

\subsection{Hund's case (a) PECs }
\label{ssec:Hcasea}

Our study relies on the PECs calculated in our previous work \cite{aymar2011}, and displayed in paper I up to the $8^{\mathrm{th}}$ dissociation limit Rb($5s \; ^2S$)$+$Sr$^+$($4d \; ^2D$), hereafter referred to the $S+D$ (or $S+D_{3/2,5/2}$) limit for simplicity (see  Appendix \ref{app:Hundsa}). Briefly, the RbSr$^+$ molecular ion is modeled as a system of two valence electrons moving in the field of ionic cores of Rb$^+$ and Sr$^{2+}$, represented by effective core potentials (ECPs). They are supplemented with core polarization potentials (CPPs) to account for core valence correlation, which depend on the orbital angular momentum of the valence electron of Rb and Sr$^+$. They are parameterized using the static polarizabilities of Rb$^+$ and Sr$^{2+}$, together with two sets of cutoff radii adjusted to recover the experimental energies of the lowest $s$, $p$ and $d$ levels in Rb and Sr$^+$. A full configuration interaction is achieved to obtain  40 Hund's case (a) PECs per symmetry in the body fixed (BF) frame (see Supplementary Material).

We have checked that our calculations yield accurate dissociation limits up to the two dissociation limits concerned in the present work, namely $S+D$ and Rb($5p \; ^2P$)$+$Sr$^+$($5s \; ^2S$), hereafter referred to the $P+S$ limit (or $P_{1/2,3/2}+S$). We note that such PECs have recently been recalculated in \cite{farjallah2025} up to the $P+S$ limit with an approach very similar to ours, and that the authors acknowledged the good agreement between their results and ours \cite{aymar2011}, as well as those of \cite{smialkowski2020} for the electronic ground state $X\, ^1\Sigma^{+}$. As mentioned in the introduction, the two colliding particles in their ground state correspond to the second dissociation limit Rb($5s \; ^2S$)$+$Sr$^+$($5s \; ^2S$), hereafter referred to the $S+S$ limit. 

As noted in paper I, the most remarkable feature is the existence of two pronounced avoided crossings involving PECs correlated to the $S+D$ and $P+S$ asymptotes:  one ({\bf X}$_{\bf 1}$) between $5\, ^1\Sigma^{+}$ and $6\, ^1\Sigma^{+}$ PECs, and the second one ({\bf X}$_{\bf 2}$) between $4\, ^3\Sigma^{+}$ and $5\, ^3\Sigma^{+}$ PECs, respectively located at $R_{X_1}=14.9$~a.u. and $R_{X_2}=13.7$~a.u. (1~a.u.=0.052917721092~nm). In paper I we predicted with a simple scattering model that they play a central role in the interpretation of FSQ and EEE inelastic processes observed experimentally. We also note the presence of a third crossing ({\bf X}$_{\bf 3}$) located at $R_{X_3}=6.8$~a.u. between $1\, ^3\Sigma^+$ and $1\, ^3\Pi$ PECs, which has been previously addressed in \cite{walewski2025b}. Such a crossing might strongly influence Feshbach resonances between ground state Rb and Sr$^+$ due to second-order SOC (see also \cite{weckesser2021b}). The crossing ({\bf X}$_{\bf 3}$) is irrelevant for the present study, but we will come back to it in the final discussion (Section \ref{sec:conclusion}).

Avoided crossings {\bf X}$_{\bf 1}$ and {\bf X}$_{\bf 2}$ manifest the presence of localized and strong non-adiabatic coupling terms (NACTs) that we do not explicitly evaluate in our electronic structure calculations. Instead, as in paper I, we linearized the PECs $5\, ^1\Sigma^{+}$ and $6\, ^1\Sigma^{+}$ around $R_{X_1}$ and the PECs $4\, ^3\Sigma^{+}$ and $5\, ^3\Sigma^{+}$ around $R_{X_2}$, and adopted a Gaussian coupling between the linearized curves
\begin{eqnarray}
    \mathrm{G}_i&=&U_{X_i} \exp[-(R-R_{X_i})^2/\delta_{i}^{2}],
    \label{eq:gaussian-cpl}
\end{eqnarray}
with i=1,2. Parameter values (in atomic units) $(U_{X_1},R_{X_1},\delta_{1})=(0.000251,14.9,0.15)$ and $(U_{X_i},R_{X_i},\delta_{i})=(0.000574,13.7,0.3)$ are chosen to retrieve the Hund's case (a) PECs after a $2 \times 2$ diagonalization.

\subsection{Extrapolation of PECs at large distances}
\label{ssec:extrapolation}

The calculated PECs $V(R)$ correlated to $S+D$ and $P+S$ and $S+S$ dissociation limits are extrapolated up to $10^4\, {\rm a.u.}$ using the standard multipolar expansion 
\begin{eqnarray}
V(R)&=&D_0+\,C_3/R^3-C_4/R^4-C_6/R^6
\label{eq:multipole}
\end{eqnarray}
where $D_0$ is the sum of atomic energy levels calculated with our method (see Supplementary Material) including terms for charge–quadrupole interaction ($C_3$), charge–induced dipole interaction ($C_4$) and van der Waals interaction ($C_6$), respectively. The values of these coefficients are listed in Table \ref{table:constants}.

For the $S+S$ and $S+D$ limits, the coefficient $C_3$ vanishes as Rb($5s$) does not have a permanent quadrupole moment. The coefficient $C_4$ is equal to half the isotropic static dipole polarizability of the Rb ground-state atom, $C_4(S+S) \equiv C_4(S+D) = \alpha_d^{(0)}(5s))/2= 159.05$~a.u.  \cite{deiglmayr2008}. The coefficient $C_6$ is then fitted in the distance range [40~a.u.;45~a.u.] containing 25 data points, to ensure the smoothest possible transition between the calculated PECs, which extend up $45$~a.u., and their long-range extrapolation. 

For the $P+S$ limit, the three coefficients are treated as fitting parameters for $\Sigma$ ($\Lambda=0$) and $\Pi$ ($\Lambda=1$) symmetries, in the same range as above. The coefficient $C_3$  is related to the electric quadrupole moment of the Rb($5p$) atom, with $C_3^{\Sigma}/C_3^{\Pi}=-2$ (see, for example, \cite{lepers2018}). The coefficient $C_4(P+S; \Lambda)$ is also anisotropic and is the result of the contributions of the isotropic and tensor dipole polarizabilities $\alpha_d^{(0)}(5p)$ and $\alpha_d^{(2)}(5p)$ of Rb($5p$) 
\begin{eqnarray}
   C_4(P+S;\Lambda)&=&0.5 \times \left[ \alpha_d^{(0)}+\alpha_d^{(2)}\frac{3\Lambda^2-L(L+1)}{L (2L-1)}  \right]
\end{eqnarray}
where $L=1$ for Rb($5p$). 

The fit led to a ratio of $C_3^{\Sigma}/C_3^{\Pi}=-2.13$ in reasonable agreement with the analytical result. The calculations reported in \cite{safronova2011} give $\alpha_d^{(0)}(5p\,^2P_{1/2})/2=407$~a.u., $\alpha_d^{(0)}(5p\,^2P_{3/2})/2=437.5$~a.u., and $\alpha_d^{(2)}(5p\,^2P_{3/2})=-167$~a.u.. This corresponds to a statistical average value of 427~a.u. for $\alpha_d^{(0)}(5p)$/2. Using the same sum-over-states formula as in \cite{safronova2011} applied to the atomic transitions with dipole moments calculated from our electronic structure method, we obtain $\alpha_d^{(0)}(5p)/2=430$~a.u. and $\alpha_d^{(2)}(5p)=-162$~a.u., so that $C_4(P+S;\Sigma)=592.5$~a.u. and $C_4(P+S;\Pi)=348.5$~a.u.. All of these values are in good agreement with each other and support the consistency of the fitting procedure and the reliability of the long-range extrapolation of the PECs, based on the quality of the atomic description in our calculations.

\begin{table}[t]
\centering
\begin{tabular}{c c c c c } 
\hline
Asymptote& State & $C_3$ & $C_4$ &$C_{6}$ \\
\hline
$S+D$ &6$^1$$\Sigma^+$ & 0.0 & 159.05 & 10138.7 \\ 
        &5$^3$$\Sigma^+$ & 0.0 & 159.05 & 10138.7 \\ 
        &4$^1$$\Pi$ & 0.0 & 159.05 & 9336.57  \\
        &4$^3$$\Pi$ & 0.0 & 159.05 & 9336.57 \\
        &2$^1$$\Delta$ & 0.0 & 159.05 & 13121.6  \\
        &2$^3$$\Delta$ & 0.0 & 159.05 & 13121.6 \\
\hline
$P+S$ &5$^1$$\Sigma^+$ & -25.4288 & 477.59& 194309.02 \\ 
        &4$^3$$\Sigma^+$ & -25.4288 & 477.59 & 194309.02 \\ 
        &3$^1$$\Pi$ & 11.9425 & 345.585 & -17451.1 \\
        &3$^3$$\Pi$ & 11.9425 & 345.585 & -17451.1 \\ 
\hline
$S+S$&2$^1$$\Sigma^+$ & 0.0  & 159.05 & 11879.357 \\ 
       &1$^3$$\Sigma^+$ & 0.0 & 159.05 & 11879.357  \\ 
\hline
\end{tabular}
\caption{Long-range coefficients (in atomic units) for the multipolar expansion of PECs correlated to the $S+D$, $S+P$, $S+S$ limits of RbSr$^+$. The $C_4$ coefficient for $S+D$ and $S+S$ is set to the value of \cite{deiglmayr2008}. The other coefficients are obtained from a fit of the Hund's case (a) PECs \cite{aymar2011}. The number of digits of the fitted coefficients are displayed for accurately reproducing the fit.}
\label{table:constants}
\end{table}

\begin{figure*}[t!]
 \centering
 \includegraphics[width=0.8\textwidth]{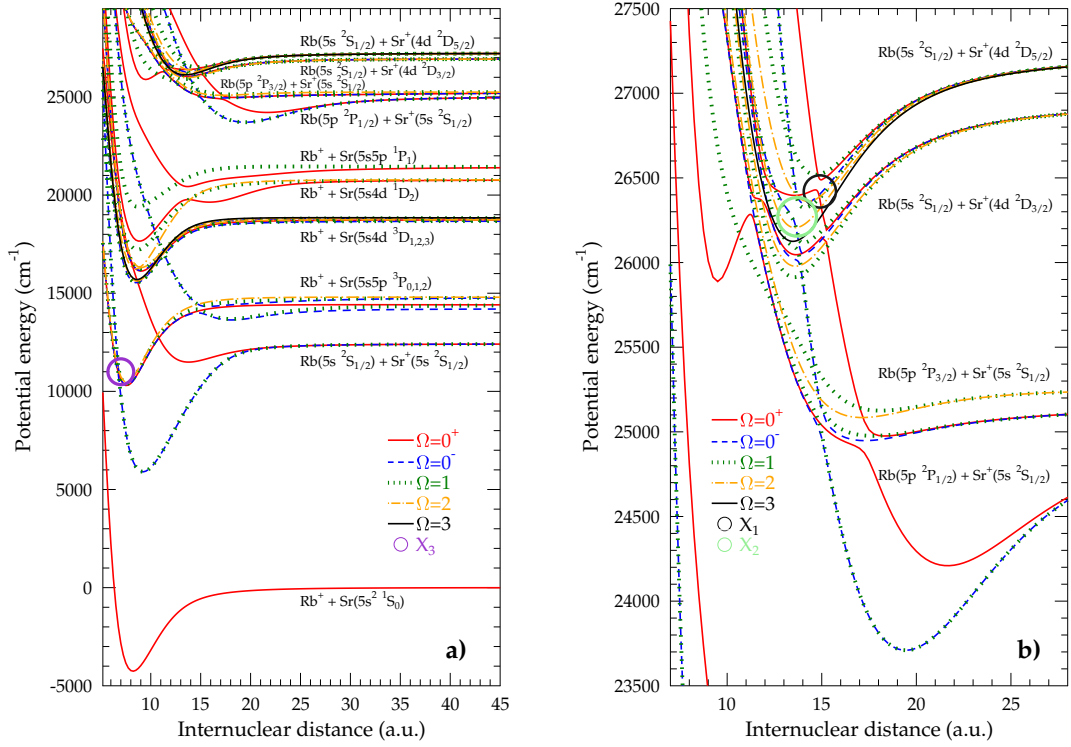}
 \caption{a) Hund's case (c) PECs of RbSr$^{+}$ in the BF frame up to the dissociation limits Rb($5s \; ^2S_{1/2}$)$+$Sr$^+$($4d \; ^2D_{3/2,5/2}$), obtained after the diagonalization of matrix $\textrm{\bf{W}}_a^{qd-so}(R)$ or $\textrm{\bf{W}}_a^{so}(R)$. The crossing {\bf X}$_{\bf 3}$ is now visible as an avoided crossing in $\Omega=0^{-},1$ symmetries. b) Blow-up of the energy region around the $S+D$ and $P+S$ dissociation limits, enlightening the complex structures generated by the avoided crossings {\bf X}$_{\bf 1}$ and {\bf X}$_{\bf 2}$ in the $\Omega=0^+$ and $\Omega=0^-$ symmetries, respectively. }
 \label{fig:PEC_C}
 \end{figure*}

\section{Spin-orbit couplings: two-step quasidiabatic model} 
\label{sec:soc-diab}

It is expected from Fig.\ref{fig:ionatomlevels} that SO interaction plays a central role in the dynamics, as also stated in paper I where we considered atomic SOC matrices of Rb and Sr$^+$, which were added to the diagonal matrix of Hund's case (a) PECs to obtain Hund's case (c) PECs after diagonalization. Following the approach originally proposed in \cite{cimiraglia1985}, and that we implemented for RbCa$^+$ \cite{xing2022} and KCs \cite{szczepkowski2022}, we derive molecular $R$-dependent SOCs from a first quasi-diabatization ($qd$) of the diagonal Hund's case (a) potential energy matrix $\textrm{\bf{W}}_a(R)$ (see Supplementary Material). 

 Briefly, at each distance $R$, the subspace of the 40 lowest Hund's case (a) eigenvectors is rotated toward a set of 40 reference states, namely the 40 lowest Hund's case (a) eigenvectors calculated at $R_{{\rm ref}}=100$~a.u., which obviously resemble to states of the separated Rb and Sr$^+$ particles. This rotation is chosen in such a way that the eight lowest rotated states match as closely as possible the eight lowest reference states. The matrix $\textrm{\bf{W}}_a(R)$ is transformed into a ''quasi-diabatic'' potential energy (non-diagonal) matrix $\textrm{\bf{W}}_a^{qd}(R)$ which is thus expressed in a basis of separated atoms, so that the atomic spin-orbit matrices $\textrm{\bf{w}}_{\Omega}^{so}$ (Appendix \ref{app:SOC}) can be added to it for each asymptote and for each value $\Omega$ of the projection of the total electronic angular momentum of the molecular ion RbSr$^+$. We included $\textrm{\bf{w}}_{\Omega}^{so}$ matrices (involving SO constants deduced from the NIST database of atomic energy levels \cite{NIST_ASD}) for the $S+D$ and $S+P$ matrices, as well as for the asymptotes Rb$^+$+Sr($5s5p\,^3P$) and Rb$^+$+Sr($5s4d\,^3D$) for completeness (see Appendix \ref{app:Hundsa}). This lead to the matrix $\textrm{\bf{W}}_a^{qd-so}(R)$ which can be used in two ways: (i) it can be diagonalized to yield the diagonal matrix $\textrm{\bf{W}}_c(R)$ of Hund's case (c) PECs (Fig. \ref{fig:PEC_C}, and Supplementary Material); (ii) it can be rotated back to the original representation of the Hund's case (a) eigenvectors, thus resulting in a non-diagonal matrix $\textrm{\bf{W}}_a^{so}(R)$ containing Hund's case (a) PECs and SO-induced shifts on the diagonal, and SOCs between them as off-diagonal elements. Note that the diagonalization of $\textrm{\bf{W}}_a^{so}(R)$ also leads to $\textrm{\bf{W}}_c(R)$.

\begin{table*}[t]
\setlength{\tabcolsep}{0.05mm}
\renewcommand{\arraystretch}{1.5} \addtolength{\tabcolsep}{3 pt}
\begin{center}
\includegraphics[width=16 cm]{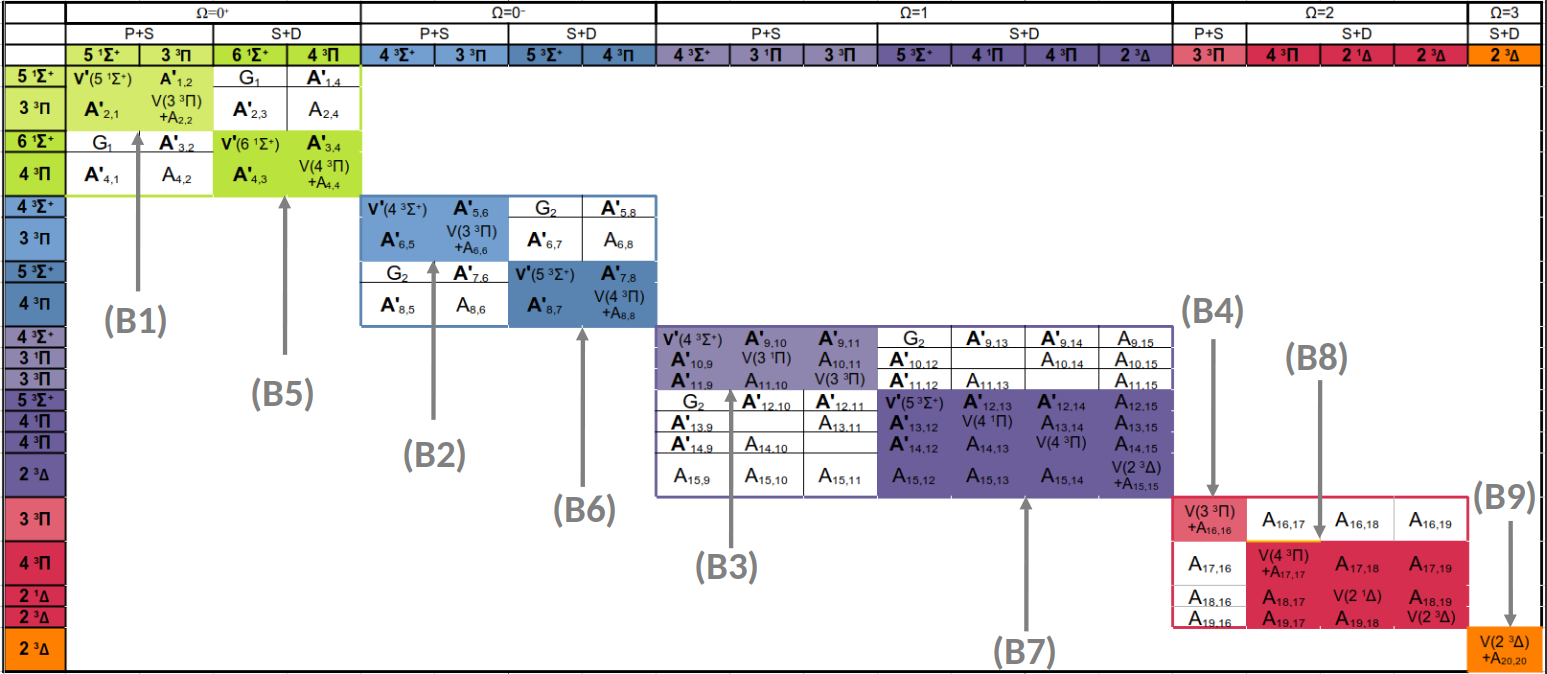}
 \end{center}
  \caption{Schematic view of the interaction sub-matrix of $\bar{\textrm{\bf{W}}}_a^{so-qd2}(R)$ restricted to the two dissociation limits $S+D$ and $P+S$ of RbSr$^+$, including all symmetries $\Omega=0^{+/-},1,2,3$ associated to the total electronic angular momentum in the body-fixed frame. Diagonal elements labeled by $V$ are Hund's case (a) PECs, while $V'$ elements refer to the locally-linearized PECs for $^1\Sigma^{+}$ and $^3\Sigma^{+}$ symmetries. The associated NACTs are labeled by G$_1$ and G$_2$ (Eq. \ref{eq:gaussian-cpl}). SOCs yielded by the first and second quasi-diabatization step are labeled by A$_{pq}$ and A'$_{pq}$, respectively, and are displayed in Fig. \ref{fig:soc_diabatized}. Direct SOCs (involving states correlated to the same asymptote) appear in colored cells, and indirect SOCs (involving states from different asymptotes) in blank cells. Empty blank cells are zero elements. Colored block matrices for each asymptote in a given symmetry are labeled with (B1) $\cdots$ (B9) notations correspond to the equations from Appendix \ref{app:SOC}.}
 \label{table:Hammatrix_1}
 \end{table*}

However, this first step is performed in the adiabatic representation of Hund's case (a), thus ignoring NACTs, in particular those associated to the avoided crossings {\bf X}$_{\bf 1}$ and {\bf X}$_{\bf 2}$. We take them into account by first restricting the $\textrm{\bf{W}}_a^{so}(R)$ matrix to the sub-matrix $\bar{\textrm{\bf{W}}}_a^{so}(R)$ containing the $S+D$ and $P+S$ asymptotes. Then we apply a second quasi-diabatization step ($qd2$) to the SOCs of $\bar{\textrm{\bf{W}}}_a^{so}(R)$ around the avoided crossings, in a similar way as the linearization of the $^1\Sigma^+$ and $^3\Sigma^+$ PEC's described above. As the SOCs matrix elements involve two different states, the global phase of each wave function must be followed with the distance through the avoided crossing to achieve the appropriate linearization. More details will be presented in a forthcoming publication. This step leads to a sub-matrix $\bar{\textrm{\bf{W}}}_a^{so-qd2}(R)$, which yields after diagonalization Hund's case (c) PECs which are indistinguishable from PECs in Fig.\ref{fig:PEC_C}.

The sub-matrix $\bar{\textrm{\bf{W}}}_a^{so-qd2}(R)$ is depicted in Table~\ref{table:Hammatrix_1}, where we collected all blocks for the symmetries $\Omega=0^{+/-},1,2,3$: this $20 \times 20$ block-diagonal matrix represents the interaction Hamiltonian matrix in the body-fixed (BF) frame of the molecular ion, which will be the input of the scattering calculations of the next section. The direct and indirect SOCs of Table \ref{table:Hammatrix_1} are shown in Fig.~\ref{fig:soc_diabatized}, with the corresponding data collected in the Supplementary Material. Indirect SOCs vanish at infinity. Direct SOCs converge at large distances towards the relevant atomic values determined by the spin–orbit splittings $\Delta E_{D} \equiv 2A_D/5 = 280.34$~cm$^{-1}$ for $S+D$ asymptote  and  $\Delta E_{P} \equiv 3A_P/2 = 237.595$~cm$^{-1}$ for $P+S$ asymptote \cite{NIST_ASD}, where $A_D$ and $A_P$ are the spin-orbit constants for Sr$^+(4d)$ and Rb$(5p)$.  Following \cite{bormotova2019} we have extrapolated our calculated SOCs curves from $45$~a.u. up to $10^4$~a.u using the functional form $\alpha+\beta_{k}/{R^{k}}$,  where $\alpha$ corresponds to the asymptotic SOC value (Appendix \ref{app:SOCfitLR}).

\begin{figure*}[t!]
 \centering
 \includegraphics[width=0.8\textwidth]{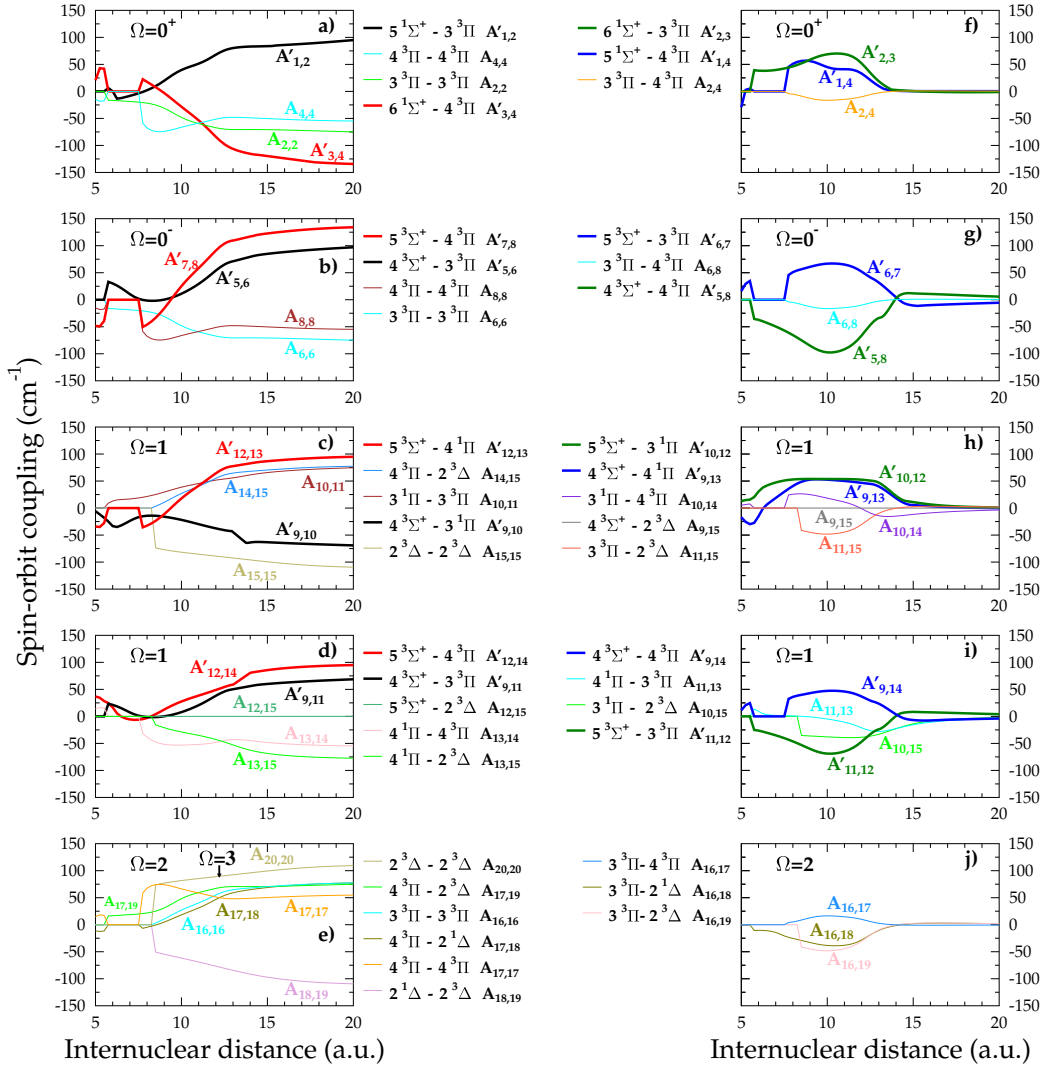}
 \caption{Direct (panels a-e) and indirect (panels f-j) SOCs A$_{pq}$ and A'$_{pq}$ as a function of the internuclear distance $R$, for the molecular symmetries $\Omega=0^+$ (a,f), $\Omega=0^-$ (b,g), $\Omega=1$ (c,d,h,i), $\Omega=2,3$ (e,j), resulting from the diabatization procedure described in the main text. Couplings which are affected by the second diabatization step (the A'$_{pq}$ elements, involving $^1\Sigma^+$ and $^3\Sigma^+$ states) are displayed with thick lines. } 
 \label{fig:soc_diabatized}
 \end{figure*}

\section{Dynamics of the R\lowercase{b}($5s$)+S\lowercase{r}$^+$($4d$) cold collision}
\label{sec:dynamic}

\begin{figure}[]
 \centering
 \includegraphics[width=\columnwidth]{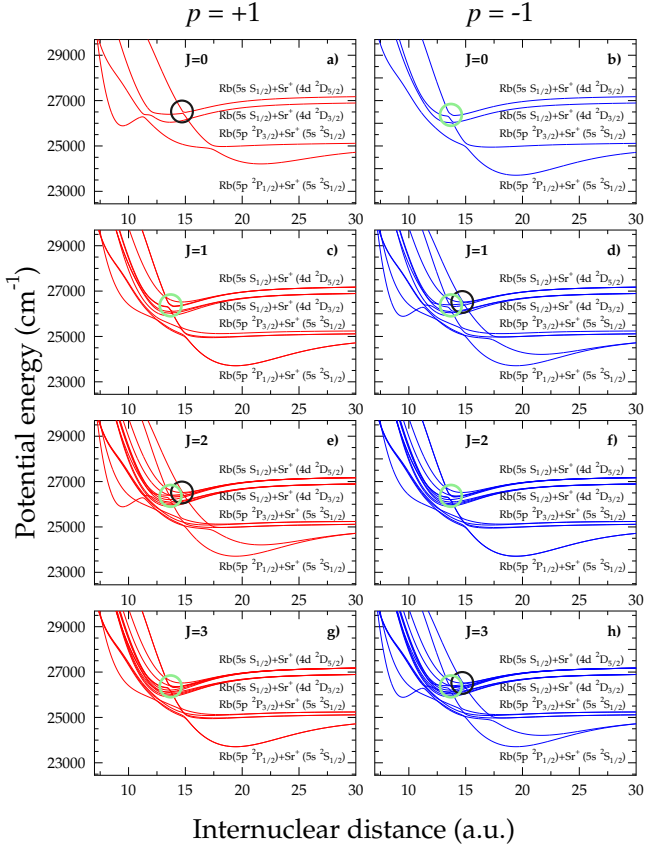}
\caption{Hund's case (e) PECs correlated to the $P+S$ and $S+D$ manifolds around crossings {\bf X}$_{\bf 1}$ and {\bf X}$_{\bf 2}$, for $+$ and $-$ parities and low values of $J$.  }
 \label{fig:pecs-e}
\end{figure}

We follow the same methodology as in our paper on the dynamics of the cold collision Li($2s$)+Ba$^+$($5d$) \cite{xing2024}. Due to the expected rotational coupling between the internal angular momenta $\vec{j}_q=\vec{L}_q+\vec{S}_q$ ($q=$ Rb, Sr$^+$) with the mutual rotation of the particles $\vec{\ell}$, the dynamics is treated in the space-fixed (SF) frame, with the interaction matrix expressed in Hund's case (e). The total angular momentum $\vec{J}=\vec{\ell}+\vec{j}_{\mathrm{Rb}}+\vec{j}_{\mathrm{Sr}^+} \equiv \vec{\ell}+\vec{j}$ with its projection $M$ on a quantization axis, and the total parity $\hat{p}$ are explicitly conserved, providing good quantum numbers $J$ and $p$. We adopted Hund's case (a)-to-(e) transformation formulated by \cite{singer1983} with the equations recalled in \cite{xing2024}. Figure \ref{fig:pecs-e} shows Hund's case (e) PECs for the total angular momenta $J\,=\, 0\, -\,3$, where the existence of {\bf X}$_{\bf 1}$ and {\bf X}$_{\bf 2}$ avoided crossings results in complex interactions between the PECs. Crossing {\bf X}$_{\bf 1}$ being linked to the $^1\Sigma^+$ symmetry, it appears only for  $p(-1)^J=+1$. For $J= 0, 1, 2$ the number of channels is 4, 11, 15, respectively. The maximum number of coupled channels is 16 for $J=3$, according to the matrix in Table \ref{table:Hammatrix_1}: depending on $J$ and $p$, $0^+$ or $0^-$ states (4 states each) are coupled to the 12 states with $\Omega \ge1$. 

A system of coupled Schrödinger equations (up to 16 equations) is generated by the total Hamiltonian matrix
\begin{equation}
\begin{aligned}
\mathbf{H}=-\dfrac{\hbar^2}{2\mu}\dfrac{d^2}{d R^2}\mathbf{I}+\dfrac{\hbar^2 \vec{\ell}^2}{2\mu R^2}\mathbf{I}+\bar{\textrm{\bf{W}}}_a^{so-qd2}(R),
\end{aligned}
\end{equation}
where $\mu$ is the reduced mass of RbSr$^+$ and $\mathbf{I}$ the identity matrix. The matrix $\mathbf{H}$ is expressed on the fully-coupled basis $(\ket{j_{\textrm{Rb}},j_{\textrm{Sr}^+})j,\ell,J,p}$. The system is solved for each value of $J$ and $p$ with log-derivative method \cite{johnson1973,alexander1987} with a constant step-size of 0.005~a.u. for 5~a.u.$<R<10^4$~a.u. and for collision energies $E_i=\hbar^2k_i^2/(2\mu)$ in the range $10^{-6}$K$<E_i/k_b<10^{-2}$K. We first extract the reaction matrix $\mathbf{K}$ holding closed and open channels and the scattering matrix $\mathbf{S}$ containing only open channels. The total cross sections are expressed as the generic sum over the partial cross sections $\sigma^{f\leftarrow i}(E_i;J,p)$
\begin{equation}
\sigma^{f\leftarrow i}(E_i) = \frac{1}{2}\sum_{p=+1,-1} \sum_{J=0}^{J_{max}}(2J+1) \sigma^{f\leftarrow i}(E_i;J,p), \\
\label{eq:xsection-tot}
\end{equation}
where the $1/2$ factor results from the average over both possible values of initial parity, and $J_{max}$ is fixed by the collision energy for which no flux can enter the scattering region. In the chosen coupled basis, the partial inelastic cross sections for a given initial state $\ket{i}=\ket{j_{\textrm{Rb}}^i,j_{\textrm{Sr}^+}^i}$, assuming that there is no initial polarization of the colliding partners, and for a final state $\ket{f}=\ket{j_{\textrm{Rb}}^f,j_{\textrm{Sr}^+}^f}$ are obtained by
\begin{equation}
\sigma^{f\leftarrow i}(E_i;J,p)= 
\dfrac{\pi}{k_i^2}  
\sum_{j_i,j_f}\,
\sum_{\substack{\ell_i,\ell_f}}
|S(j_f\, \ell_f \gets j_i\, \ell_i;J,p)|^2,
\label{eq:xsection-Jp}
\end{equation}
where the sums over $j_i$ and $j_f$ correspond to all vectors generated by the chosen values of $(j_{\textrm{Rb}}^i,j_{\textrm{Sr}^+}^i)$ and $(j_{\textrm{Rb}}^f,j_{\textrm{Sr}^+}^f)$.

\begin{figure}[h]
\centering
\includegraphics[width=\columnwidth]{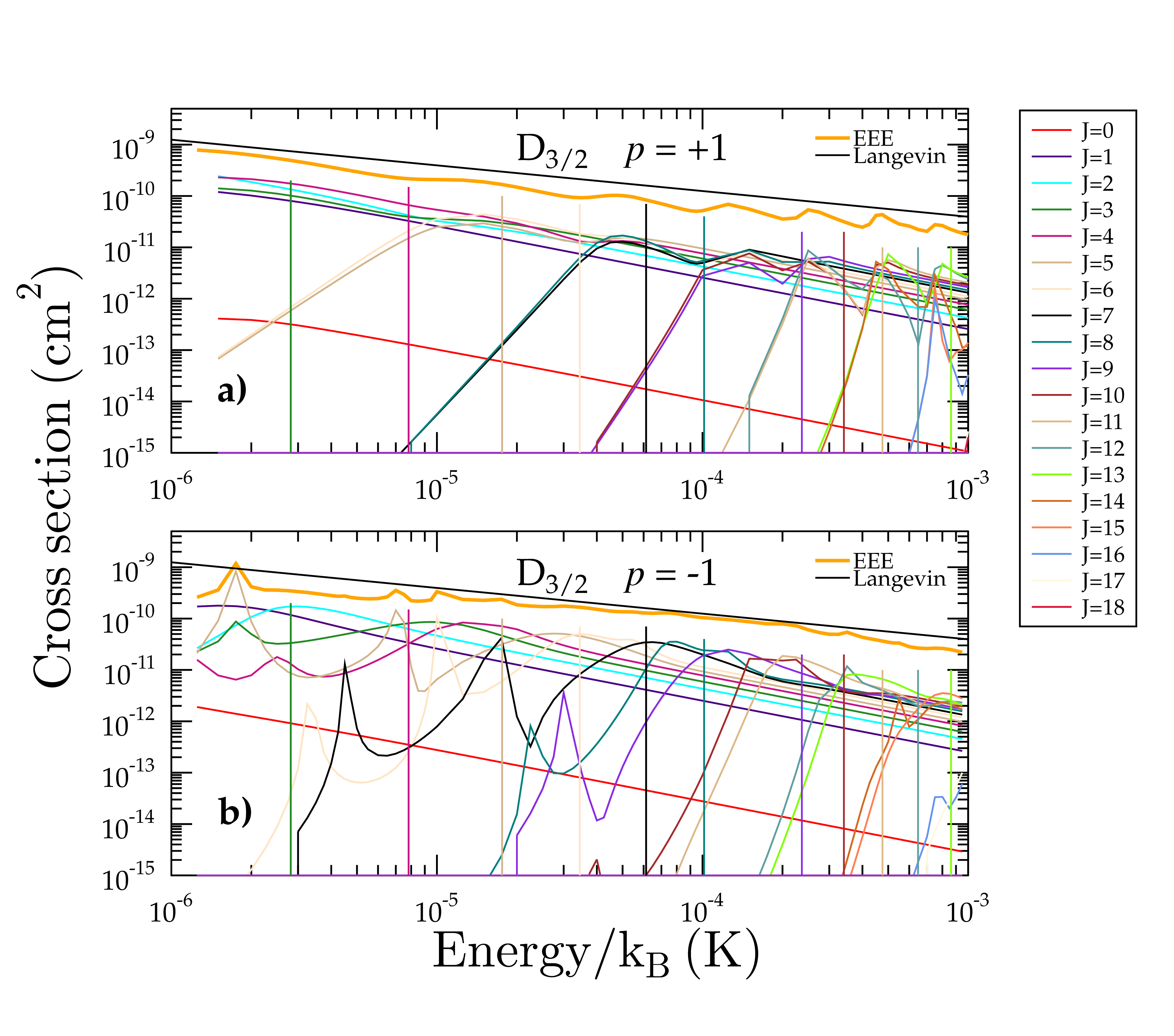}
\caption{Partial cross sections $\sigma^{f\leftarrow i}(E_i;J,p)$ as a function of collision energy (expressed in K) for EEE starting from the entrance channel $S+D_{3/2}$. The total cross section of the EEE process (upper yellow thick solid line) and the Langevin cross section (upper black solid line) are also displayed. The vertical colored lines locate the heights of the centrifugal barriers for $\ell>2$, the ones for $\ell \le 2$ lying below $E/k_b \approx 10^{-6}$~K. They indicate around which energy the reactions become barrierless for the corresponding $J$ value.}
\label{fig:xjtp_32_eee_dense}
\end{figure}

An example of partial inelastic cross sections (Eq. \ref{eq:xsection-Jp}) is displayed in Fig. \ref{fig:xjtp_32_eee_dense} for EEE process (Eq. \ref{eq:EEE}) starting from $S+D_{3/2}$. The upper bound represented by the classical Langevin cross section $\sigma_L=2\pi [C_4(S+D)]^{1/2}E^{-1/2}_i$ is also drawn. The graph reveals that entrance channels with $J=0$ to 5 are already open at $E/k_b \approx 10^{-6}$~K, and the undulations of the computed total cross section are generated by opening consecutive $J$ channels related to partial waves $\ell$ as the collision energy increases. For $p=-1$, sharp structures appear in $\sigma^{f\leftarrow i}(E_i;J,p)$ due to scattering resonances in the entrance channels, some of them remaining visible in the total cross sections.

\section{Branching ratios of the R\lowercase{b}($5s$)+S\lowercase{r}$^+$($4d$) cold collision}
\label{sec:branching}

\begin{figure}[h!]
\centering
\includegraphics[width=\columnwidth]{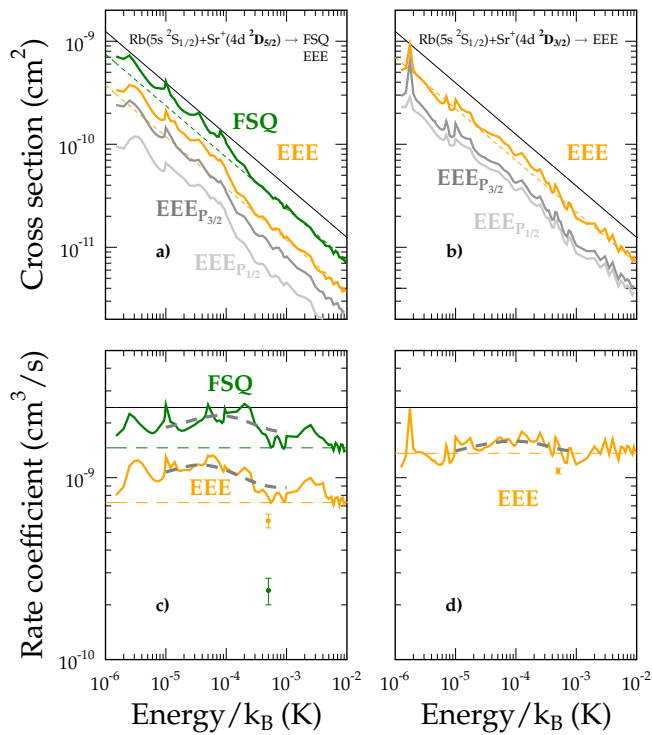}
\caption{Computed total inelastic cross sections (panels a) and b)) and rate coefficients (panels c) and d)) for both incoming channels $S+D_{3/2,5/2}$. Green solid lines: FSQ. EEE: orange solid lines. Langevin cross section and rate: black solid line. Scaled Langevin quantities (see text): colored dashed lines. In panels a) and b): EEE toward $P_{1/2}+S$ (light grey lines), EEE toward $P_{3/2}+S$ (dark grey lines). In panels c) and d): thermalized rate coefficients (gray dashed lines). The green dot and yellow square with error bars represent the measured data for FSQ and EEE.}
\label{fig:xsections_rates_gray}
\end{figure}

The computed total cross sections (Eq. \ref{eq:xsection-tot}) $\sigma^{f\leftarrow i}(E_i)$ of the EEE and FSQ processes and related rate coefficients $K^{f\leftarrow i}(E_i)= \sqrt{2E_i/\mu}\, \sigma^{f\leftarrow i}(E_i)$ are displayed in Fig. \ref{fig:xsections_rates_gray} for both entrance channels. They are compared with the upper limits of the Langevin model, namely $\sigma_L$ and $K_L=\sqrt{2E_i/\mu}\, \sigma_L=2.43 \times 10^{-9}$~cm$^3$/s. To obtain rate coefficients that can be compared to the measured one, we calculated the thermalized rate coefficients $K^{f\leftarrow i}_{\text{th}}(T_{\text{eff}})$ assuming a Maxwell-Boltzmann distribution of collisional energies for a given effective temperature $T_{\text{eff}}$,
\begin{equation}
    T_{\mathrm{eff}}=\frac{\mu}{m_{\mathrm{Sr^+}}}T_{\mathrm{Sr^+}}+\frac{\mu}{m_{\mathrm{Rb}}}T_{\mathrm{Rb}}.
\end{equation}
In principle, the effective temperature $T_{\mathrm{eff}}$ in the center-of-mass  ($m_{Sr^+}$ and $m_{Rb}$ are the masses of the ion and the atom) might be determined by the individual  temperatures $T_{Sr^+} \approx 40\, \mu$K and $T_{Rb}\approx 3\, \mu$K, resulting in $T_{\mathrm{eff}} \approx 21\,\mu$K in the experiment \cite{benshlomi2020}. However, the value $T_{\mathrm{eff}} \approx 0.5$~mK was reported in paper I, well above the $s$-wave limit around 80~nK, which presents a significant challenge for experiments. However, it is worth recalling that the experimental velocity distribution is actually very different from a Maxwell-Boltzmann distribution (see, for instance, \cite{hall2013b}).

The computed total cross sections reflect the expected variation in $1/\sqrt{E_i}$ of the Langevin model, with a significant reduction with respect to the Langevin cross section induced by quantum effects. We fit $\sigma^{f\leftarrow i}(E_i)$ between 1~mK and 10~mK with a Langevin-type expression, producing $\sigma^{f\leftarrow i}(E_i)/\sigma_L=0.60$ for FSQ, and 0.30 (resp. 0.56) for EEE starting from $S+D_{5/2}$ (resp. $S+D_{3/2}$). The total inelastic cross section (EEE+FSQ) starting from $S+D_{5/2}$ is predicted close to the classical Langevin limit. The FSQ cross section is twice that of the EEE cross section when the Sr$^+$ ion is prepared in the $^2D_{5/2}$ state, as in our previous study on Li/Ba$^+$ \cite{xing2024}. Partial cross sections for EEE toward the $P_{1/2}+S$ and $P_{3/2}+S$ exit channels are also displayed for completeness: as expected, the latter is larger than the former, as $P_{3/2}+S$ is the closest exit channel in energy. As noted above, the cross sections exhibit narrow structures induced by scattering resonances in the entrance channel. 

These resonances are still visible in rate coefficients, but are washed out in the thermalized rates, so that it is likely that they could not be detected in an experiment. In paper I, measured probability values at $T_{\text{eff}}=0.5$~mK for each process were reported, which we compare with the computed probabilities ${\cal P}^{3/2,5/2}_{EEE,FSQ}=K^{f\leftarrow i}(E_i)/K_L$ in Table \ref{tab:results} (where the upper index refers to the  value of $j^{Sr^+}$ in the entrance channel). In Fig. \ref{fig:xsections_rates_gray} the measured probabilities of paper I are transformed in rate coefficients by multiplying them by $K_L$. The measured total probability ${\cal P}^{5/2}_{EEE}+{\cal P}^{5/2}_{FSQ}=0.34$ is almost three times less than our predicted value (0.96). The calculated rate coefficient for the FSQ process overestimates the experimental observations by slightly less than one order of magnitude, while an agreement is obtained within a factor of 1.5 for the EEE process. A consistent pattern has previously been found for collisions between ground-state Li atoms and metastable Ba$^+$ ions \cite{xing2024} collisions. This discrepancy is probably due to the absence of the Zeeman sublevels of the particles in our approach. Indeed, in paper I, both particles were polarized: $^{87}$Rb($5s\,\,^2S_{1/2}$) in the hyperfine Zeeman level $\ket{f_{Rb}=2,\, m_f=-2}$  implying an electronic spin projection $m_s=-1/2$, and $^{88}$Sr$^+$($4d\,\,^2D_{5/2}$) (with  zero nuclear spin) in the $m_j=\pm5/2$ level. The measured rates were found identical for $m_j=\pm5/2$, thus independent of the relative orientation of the magnetic moments.  In contrast, $^{88}$Sr$^+$ ions in the ($4d\,^2D_{3/2}$) state were not polarized.

The satisfactory agreement of our calculations with the measurements for EEE with unpolarized $^{88}$Sr$^+$ ($4d\,^2D_{3/2}$) ions as well as polarized $^{88}$Sr$^+$ ($4d\,^2D_{5/2}$) ions suggests that the dynamics of EEE is weakly dependent on the initial Zeeman sublevels. This might be consistent with Hund's case (e) PECs (Fig. \ref{fig:pecs-e}) showing that the entrance flux has to travel through multiple crossings where the memory of  initial Zeeman sublevels would be lost. In contrast, FSQ dynamics relies on a very specific avoided crossing that we expect to strongly depend on initial Zeeman sublevels. The extension of our dynamical model is in progress.

\begin{table}[h!]\footnotesize
\setlength{\tabcolsep}{0.05mm}
\renewcommand{\arraystretch}{1.6} \addtolength{\tabcolsep}{3 pt}
\begin{center}
    \begin{tabular}{lcc} \hline
Probability                  &Experiment               &This work          \\
                             &                         &$K^{f\leftarrow i}(E_i)/K_L$ \\ 
                             \hline      
${\cal P}_{EEE}^{3/2}$              & $\mathbf{0.45\pm 0.018}$&0.64  \\
${\cal P}_{EEE+FSQ}^{5/2}$          & $\mathbf{0.34\pm 0.017}$&0.96 \\
${\cal P}_{EEE}^{5/2}/P_{EEE}^{3/2}$& $\mathbf{0.53\pm 0.043}$& 0.50 \\
${\cal P}_{FSQ}^{5/2}/P_{EEE}^{5/2}$& $\mathbf{0.39\pm 0.050}$& 2.05 \\
${\cal P}_{FSQ}^{5/2}$              & $0.10\pm 0.01 $& 0.66  \\
${\cal P}_{EEE}^{5/2}$              & $0.24 \pm 0.02 $&  0.32 \\ \hline
    \end{tabular}
    \end{center}
    \caption{
    Measured \cite{benshlomi2020} and calculated probabilities ${\cal P}^{3/2,5/2}_{EEE,FSQ}=K^{f\leftarrow i}(E_i)/K_L$ for EEE and FSQ processes, starting from the two incoming channels labeled with the initial value of $j^{Sr^+}$, at the reported experimental collisional energy $E_i/k_B=0.5$~mK. The first 4 lines (in boldface) were reported in paper I, while the two last lines are deduced from the previous one for the present work. The used Langevin rate coefficient is $K_{L}=2.43\times 10^{-9}$ cm$^3/$s considering $C_4=159.05$ a.u. \cite{deiglmayr2008} and the reduced mass $\mu=79\,664.38$ a.u..} 
    \label{tab:results}
\end{table}

\section{Concluding remarks and perspectives}
\label{sec:conclusion}

In this work, we described the quantum dynamics of cold collisions between a ground state atom and a metastable excited ion, with the specific example of Rb and Sr$^+$ which followed our previous work on Li and Ba$^+$ \cite{xing2024}. Despite the apparent simplicity of such diatomic systems, our work highlights the complexity of the dynamics induced by non-adiabatic coupling terms, for which a specific theoretical treatment based on original quasi-diabatic models had to be developed. Starting from our calculated Hund's case (a) potential energy curves where molecular spin-orbit interaction has been introduced via a quasi-diabatic transformation, we calculated cross section and reaction rates for two processes, namely electronic excitation exchange (EEE) and fine-structure quenching (FSQ), while no charge exchange is expected to happen. Our results are found to be in satisfactory agreement with the experiment for the EEE process, just as was the case for Li-Ba$^+$ collisions, confirming the quality of our molecular data not only for the excited electronic manifold, but also with respect to the modeling of spin-orbit interaction. In contrast, a significant discrepancy in the FSQ rates is obtained, which is assigned to a presumably pronounced dependence of the process on the initial Zeeman sublevels in which the reactants are prepared in the experiment. This represents a natural extension of our formalism by including the corresponding quantum numbers in the basis set chosen for the dynamical calculations.

\begin{figure}[h!]
\centering
\includegraphics[width=\columnwidth]{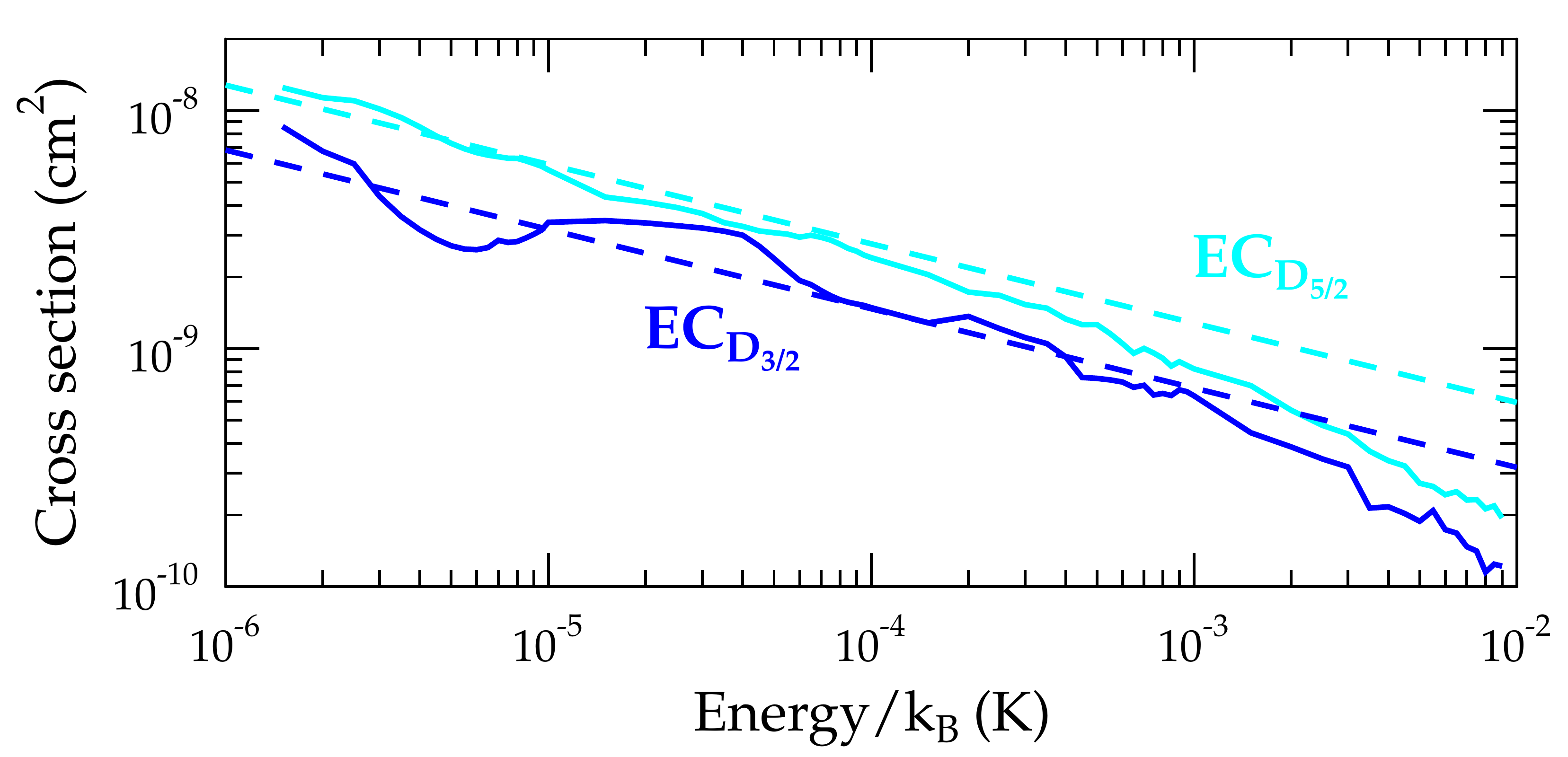}
\caption{Computed elastic cross sections for collisions between ground state Rb and excited metastable Sr$^+$. Blue (cyan) solid lines: $S+D_{3/2}$ ($S+D_{5/2}$) entrance channel. Dashed lines: corresponding fit with the inverse power law $\sigma(E_i) =\xi \, (E_i/k_B)^{-1/3}$ following \cite{cote2000a} for $10^{-6}$~K$<E_i/k_b<10^{-3}$~K. $S+D_{3/2}$ ($S+D_{5/2}$): $\xi=6.83 \times 10^{-11}$~K$^{1/3}$cm$^2$ ($\xi=1.28 \times 10^{-10}$~K$^{1/3}$cm$^2$). }
 \label{fig:xjtp_elastic5232}
 \end{figure}

These results, which assess the quality of our molecular data and of their implementation through quasi-diabatization, are encouraging to explore other scattering properties of the system. For instance, we can calculate the partial cross section (leading to a total cross section $\sigma(E_i)$ with Eq. \ref{eq:xsection-tot}) for such an elastic collision (EC) involving a metastable excited particle,
\begin{equation}
\sigma^{i}(E_i;J,p)= 
\dfrac{\pi}{k_i^2}  
\sum_{j_i,j_i^{'}}\,
\sum_{\ell_i,\ell_i^{'}}
|\delta_{j_i,j_i^{'}}\,\delta_{\ell_i,\ell_i^{'}}\,
- S(j'_i\, \ell'_i \gets j_i\, \ell_i;J,p)|^2,
\label{eq:xsection-Jp-elastic_OA}
\end{equation}
where the sum runs over all allowed values of $j_i=|j^{\rm{Rb}}-j^{{\rm Sr}^+}|,\,\dots ,\,j^{\rm{Rb}}+j^{{\rm Sr}^+}$ and $\ell$. The case of RbSr$^+$ is appropriate to test the semiclassical model of \cite{cote2000a} of the cross section in the regime of numerous partial waves
\begin{equation}
\sigma(E_i)=\pi \left(\frac{\mu C_4^2}{\hbar^2}\right)^{1/3} \left(1+\frac{\pi^2}{16}\right) E_i^{-1/3} \equiv \xi \, (E_i/k_B)^{-1/3}.
\label{eq:EC-cote}    
\end{equation}
There are already several contributing partial waves for $T=10^{-6}$~K, and we see in Fig. \ref{fig:xjtp_elastic5232} that this behavior is valid up to $\approx T=10^{-3}$~K. It is worth noticing that the model of \cite{cote2000a} was formulated for elastic collisions between ground state Na and Na$^+$, characterized by two simple and non-interacting Hund's case (a) PECs. Applying Eq. \ref{eq:EC-cote} in the case of RbSr$^+$, we find $\xi=1.23 \times 10^{-11}$~K$^{1/3}$cm$^2$, while a fit of the EC cross sections gives $\xi_{3/2}=6.83 \times 10^{-11}$~K$^{1/3}$cm$^2$ and $\xi_{5/2}=1.28 \times 10^{-10}$~K$^{1/3}$cm$^2$. These large values are assigned to the larger effective size of the excited Sr$^+$ ion, inducing larger elastic cross sections.

\begin{figure}[h!]
\centering
\includegraphics[width=1\columnwidth]{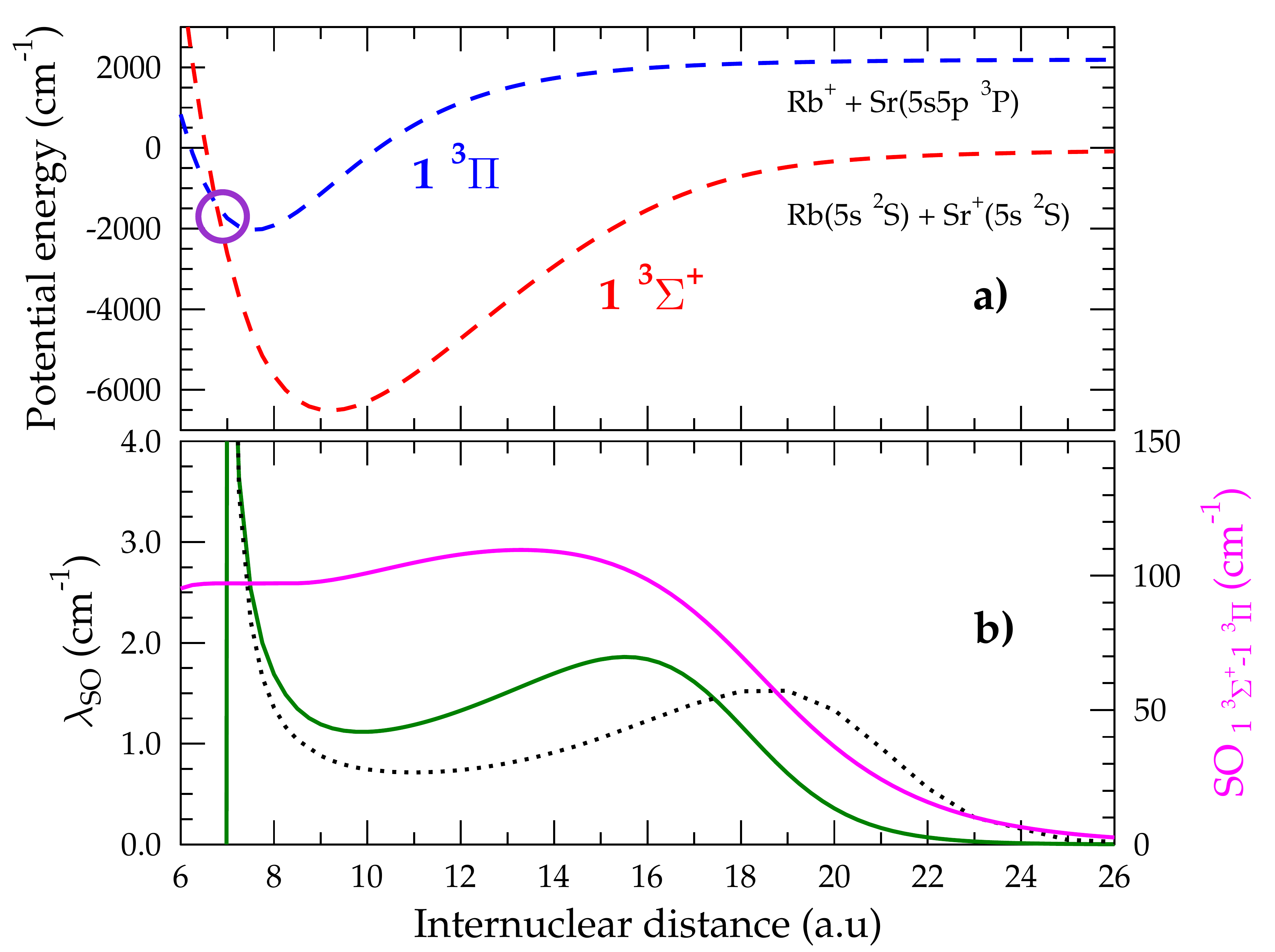}
\caption{a) Hund's case (a) PECs $1\,^3\Sigma^+$ and $1\,^3\Pi$ showing their crossing {\bf X}$_{\bf 3}$ highlighted with a circle at 6.8~a.u.. b) Computed indirect SOC matrix element $A_{X_3}(R)$ between these states (magenta line, right vertical scale), and second-order SO coefficient $\lambda_{\textrm{SO}}(R)$ (green line, left vertical scale) compared to the unscaled calculation from \cite{walewski2025b} (black dotted line).}
 \label{fig:X3}
 \end{figure}

The present quasi-diabatic approach can also be tested for the crossing {\bf X}$_{\bf 3}$ mentioned in Section \ref{sec:PECSBF}, which occurs at $R=6.8$~a.u. between the $1\,^3\Sigma^+$ PEC correlated to $S+S$ and the $1\,^3\Pi$ PEC correlated to the next dissociation limit Rb$^+$+Sr($5s5p\, ^3P$). An indirect SOC denoted $A_{X_3}(R)$ exists between these states that we extracted from the matrix $\textrm{\bf{W}}_a^{so}(R)$ (Fig. \ref{fig:X3}). Following \cite{walewski2025b}, one can derive a second-order SO coefficient $\lambda_{\textrm{SO}}(R)$ 
\begin{equation}
    \lambda_{\textrm{SO}}(R)=\frac{2}{3}\frac{(A_{X_3}(R))^2}{V_{1\,^3\Pi}(R)-V_{1\,^3\Sigma^+}(R)}
\label{eq:SOC-X3}
\end{equation}
based on a second-order perturbation approach, which is represented in Fig.\ref{fig:X3}. Although the value of $A_{X_3}(R)$ is not provided in \cite{walewski2025b}, our determination $\lambda_{\textrm{SO}}(R)$ is in good agreement with that of \cite{walewski2025b}, however, with a slight shift at large distances which may be due to small differences in Hund's case (a) PECs. In view of these results, the present quasi-diabatization approach appears as a promising way to take into account spin-orbit effects in highly excited molecular states.

\section*{Acknowledgments}
This work was supported by CNRS International Emerging Action (IEA) - ELKH, 2023-2024; Program Hubert Curien ”BALATON” (Campus France, GrantNo. 49848TC); NKFIH TÉT-FR, 2023-2024 (2021-1.2.4- TÉT-2022-00069). 
A.O. thanks the Campus France Excellence Hongrie program and the NKFIH-2025-1.2.4-TÉT-2025-00061 project for their support.
T.J. gratefully acknowledges financial support '' Bourse du Gouvernement Français'' and the Hungarian State Doctoral Scholarship.

\clearpage
\appendix
\section{Hund's case (a) PECs for RbSr$^+$}
\label{app:Hundsa}
The Hund's case (a) PECs of RbSr$^+$, already displayed in paper I, are recalled here on a larger scale, with a blow-up of the region of the $P+S$ and $S+D$ dissociation limits, probing that the main dynamical effects initially arise from avoided crossings in the Hund's case (a) PEC's. The corresponding data in atomic units of energy are provided in the Supplementary Material and are transformed in cm$^{-1}$ considering the atomic energies calculated with our approach (Table \ref{tab:atomic-levels}).

\begin{table}[h!]
\begin{center}
\begin{tabular}{|c|c|} \hline
Level& Energy (a.u.) \\ \hline     
Rb($5s$)       & -0.153507732592 \\
Rb($5p$)       & -0.095471900952 \\
Sr$^+$($5s$)   & -0.405204475915 \\
Sr$^+$($4d$)   & -0.338269312485 \\
Sr($5s^2\,^1S$)&-0.615276171121 \\
Sr($5s5p\,^3P$)&-0.548793716582 \\
Sr($5s4d\,^3D$)&-0.529764040902 \\
Sr($5s4d\,^1D$)&-0.520635929285 \\
Sr($5s5p\,^1P$)&-0.517653559858 \\ \hline
\end{tabular}
\end{center}
\caption{Atomic energy levels calculated with the present approach.}
\label{tab:atomic-levels}
\end{table}

\begin{figure*}
 \centering
 \includegraphics[width=\textwidth]{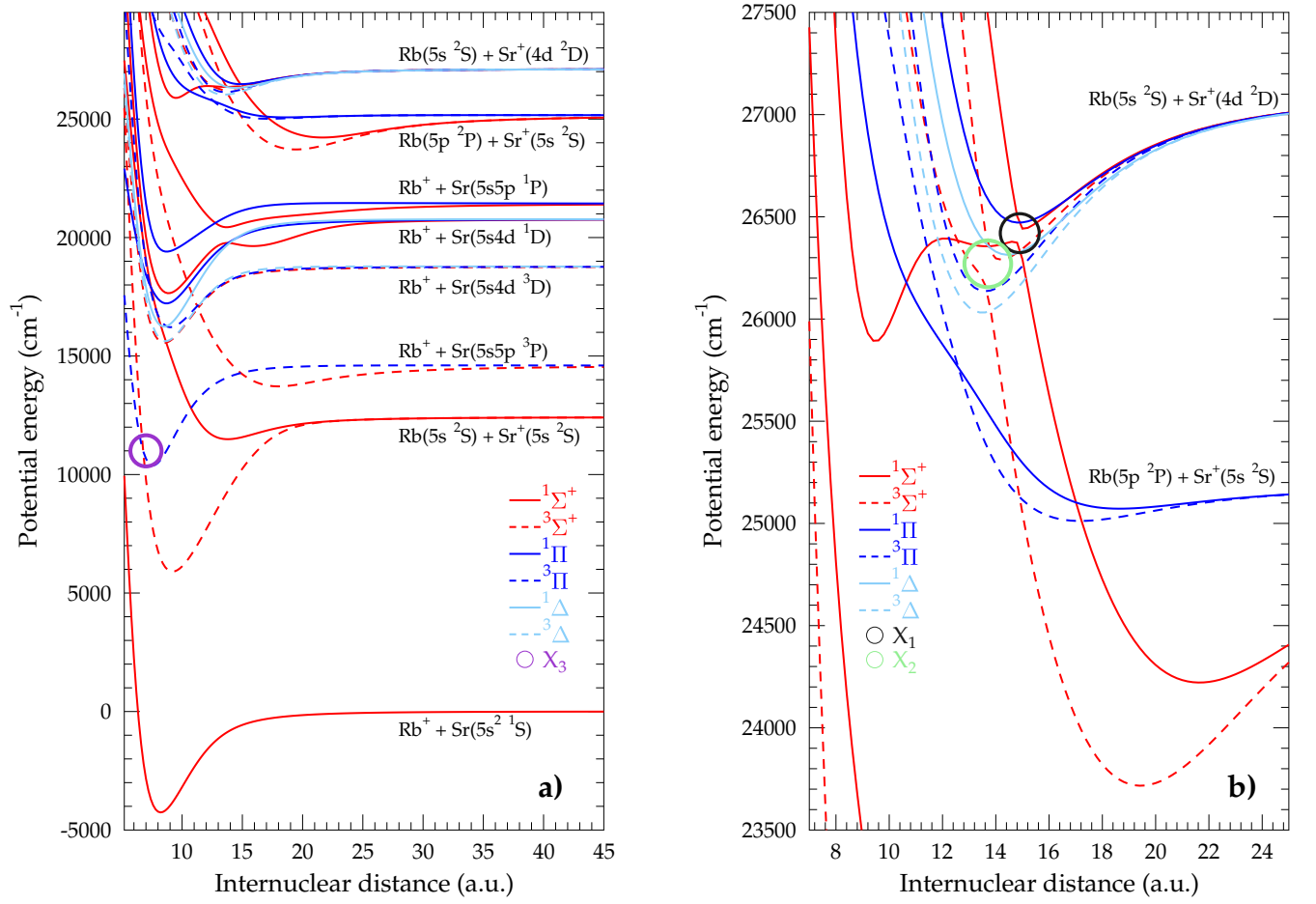}
 \caption{{a) Hund's case (a) PECs of the RbSr$^{+}$  molecule up to the eighth dissociation limit Rb($5s \; ^2S$)$+$Sr$^+$($4d \; ^2D$). The {\bf X}$_{\bf 3}$ crossing is highlighted with a circle. b) Zoom on the energy region of the $P+S$ and $S+D$ dissociation limits, where the two avoided crossings {\bf X}$_{\bf 1}$ and {\bf X}$_{\bf 2}$ are highlighted with circles. }}
 \label{fig:PEC_A}
 \end{figure*}
\clearpage
\section{Atomic spin-orbit matrices for the $P+S$ and $S+D$ dissociation limits}
\label{app:SOC}

We recall the atomic spin-orbit matrices $\textrm{\bf{w}}_{\Omega}^{so}$ written in the separated atom basis, already given in paper I, with differences of sign for several matrix elements as a result of the well-defined phase-convention adopted in the present work for the frame transformation and the quasi-diabatic approach. 

\subsection{Asymptote $S+P$}
\begin{equation}
\begin{array}{c@{\hspace{0.5em}}c@{\hspace{0.5em}}c}
&
\begin{array}{cc}
5\,^1\Sigma^+ & 3\,^3\Pi^+
\end{array}
&
\\
\textrm{\bf{w}}_{\Omega=0^+}^{so}=
&
\left(
\begin{array}{cc}
0 & \dfrac{A_{P}}{\sqrt{2}} \\
\dfrac{A_{P}}{\sqrt{2}} & -\dfrac{A_{P}}{2}
\end{array}
\right)
&
\begin{array}{c}
5\,^1\Sigma^+ \\
3\,^3\Pi^+
\end{array}
\end{array}
\label{eq:P+S_so_0+}
\end{equation}

\begin{equation}
\begin{array}{c@{\hspace{0.5em}}c@{\hspace{0.5em}}c}
&
\begin{array}{cc}
4\,^3\Sigma^+ & 3\,^3\Pi^+
\end{array}
&
\\
\textrm{\bf{w}}_{\Omega=0^-}^{so}=
&
\left(
\begin{array}{cc}
0 & \dfrac{A_{P}}{\sqrt{2}} \\
\dfrac{A_{P}}{\sqrt{2}} & -\dfrac{A_{P}}{2}
\end{array}
\right)
&
\begin{array}{c}
4\,^3\Sigma^+ \\
3\,^3\Pi^+
\end{array}
\end{array}
\label{eq:P+S_so_0-}
\end{equation}

\begin{equation}
\begin{array}{c@{\hspace{0.5em}}c@{\hspace{0.5em}}c}
&
\begin{array}{ccc}
4\,^3\Sigma^+ & 3\,^1\Pi & 3\,^3\Pi
\end{array}
&
\\
\textrm{\bf{w}}_{\Omega=1}^{so}=
&
\left(
\begin{array}{ccc}
0 & -\dfrac{A_{P}}{2} & \dfrac{A_{P}}{2} \\
-\dfrac{A_{P}}{2} & 0 & \dfrac{A_{P}}{2} \\
\dfrac{A_{P}}{2} & \dfrac{A_{P}}{2} & 0
\end{array}
\right)
&
\begin{array}{c}
4\,^3\Sigma^+ \\
3\,^1\Pi \\
3\,^3\Pi
\end{array}
\end{array}
\label{eq:P+S_so_1}
\end{equation}

\begin{equation}
\begin{array}{c@{\hspace{0.5em}}c@{\hspace{0.5em}}c}
&
\begin{array}{c}
3\,^3\Pi
\end{array}
&
\\
\textrm{\bf{w}}_{\Omega=2}^{so}=
&
\left(
\begin{array}{c}
\dfrac{A_{P}}{2}
\end{array}
\right)
&
\begin{array}{c}
3\,^3\Pi
\end{array}
\end{array}
\label{eq:P+S_so_2}
\end{equation}

\subsection{Asymptote $S+D$} 
\begin{equation}
\begin{array}{c@{\hspace{0.5em}}c@{\hspace{0.5em}}c}
&
\begin{array}{cc}
6\,^1\Sigma^+ & 4\,^3\Pi
\end{array}
&
\\
\textrm{\bf{w}}_{\Omega=0^+}^{so}=
&
\left(
\begin{array}{cc}
0 & -A_{D}\sqrt{\frac{3}{2}} \\
-A_{D}\sqrt{\frac{3}{2}} & -\dfrac{A_{D}}{2}
\end{array}
\right)
&
\begin{array}{c}
6\,^1\Sigma^+ \\
4\,^3\Pi
\end{array}
\end{array}
\label{eq:S+D_so_0+}
\end{equation}

\begin{equation}
\begin{array}{c@{\hspace{0.5em}}c@{\hspace{0.5em}}c}
&
\begin{array}{cc}
5\,^3\Sigma^+ & 4\,^3\Pi
\end{array}
&
\\
\textrm{\bf{w}}_{\Omega=0^-}^{so}=
&
\left(
\begin{array}{cc}
0 & A_{D}\sqrt{\frac{3}{2}} \\
A_{D}\sqrt{\frac{3}{2}} & -\dfrac{A_{D}}{2}
\end{array}
\right)
&
\begin{array}{c}
5\,^3\Sigma^+ \\
4\,^3\Pi
\end{array}
\end{array}
\label{eq:S+D_so_0-}
\end{equation}

\begin{equation}
\begin{array}{c@{\hspace{0.5em}}c@{\hspace{0.5em}}c}
&
\begin{array}{cccc}
5\,^3\Sigma^+ & 4\,^1\Pi & 4\,^3\Pi & 2\,^3\Delta
\end{array}
&
\\
\textrm{\bf{w}}_{\Omega=1}^{so}=
&
\left(
\begin{array}{cccc}
0 & A_{D}\frac{\sqrt{3}}{2} & A_{D}\frac{\sqrt{3}}{2} & 0 \\
A_{D}\frac{\sqrt{3}}{2} & 0 & -\frac{A_{D}}{2} & -\frac{A_{D}}{\sqrt{2}} \\
A_{D}\frac{\sqrt{3}}{2} & -\frac{A_{D}}{2} & 0 & \frac{A_{D}}{\sqrt{2}} \\
0 & -\frac{A_{D}}{\sqrt{2}} & \frac{A_{D}}{\sqrt{2}} & -A_{D}
\end{array}
\right)
&
\begin{array}{c}
5\,^3\Sigma^+ \\
4\,^1\Pi \\
4\,^3\Pi \\
2\,^3\Delta
\end{array}
\end{array}
\label{eq:S+D_so_1}
\end{equation}

\begin{equation}
\begin{array}{c@{\hspace{0.5em}}c@{\hspace{0.5em}}c}
&
\begin{array}{ccc}
4\,^3\Pi & 2\,^1\Delta & 2\,^3\Delta
\end{array}
&
\\
\textrm{\bf{w}}_{\Omega=2}^{so}=
&
\left(
\begin{array}{ccc}
\frac{A_{D}}{2} & \frac{A_{D}}{\sqrt{2}} & \frac{A_{D}}{\sqrt{2}} \\
\frac{A_{D}}{\sqrt{2}} & 0 & -A_{D} \\
\frac{A_{D}}{\sqrt{2}} & -A_{D} & 0
\end{array}
\right)
&
\begin{array}{c}
4\,^3\Pi \\
2\,^1\Delta \\
2\,^3\Delta
\end{array}
\end{array}
\label{eq:S+D_so_2}
\end{equation}

\begin{equation}
\begin{array}{c@{\hspace{0.5em}}c@{\hspace{0.5em}}c}
&
\begin{array}{c}
2\,^3\Delta
\end{array}
&
\\
\textrm{\bf{w}}_{\Omega=3}^{so}=
&
\left(
\begin{array}{c}
A_{D}
\end{array}
\right)
&
\begin{array}{c}
2\,^3\Delta
\end{array}
\end{array}
\label{eq:S+D_so_3}
\end{equation}

\clearpage
\section{Extrapolation of the SOCs at long-range}
\label{app:SOCfitLR}
\renewcommand{\arraystretch}{1.6} \addtolength{\tabcolsep}{3 pt}

Following the long range behavior previously proposed for heteronuclear alkali-dimers in ref. \cite{bormotova2019} we have extrapolated our calculated SOCs from $45$~a.u. up to $10^4$~a.u  using the functional form $\alpha+\beta_{k}/R^{k}$  where  $\alpha$ corresponds to the asymptotic (atomic) SOC value. The parameter $\beta$ is obtained by fitting the calculated SOCs in the $43-45$~a.u range. A smooth match to long-range behavior is achieved with $k=6$ for direct SOCs (Table~\ref{table:directSOC_P+S}, ~\ref{table:directSOC_S+D}) and $k=3$ for indirect SOCs (Table~\ref{table:indirectSOC}).

\begin{table}[h!]
\centering
\begin{tabular}{c c c c c  } 
\hline
limit& $\Omega$ & $A_{i,j}$ & $\alpha$ & $\beta_{6}$ (a.u) \\ \hline
$P+S$ & $\Omega=0^{+}$  & $A_{1,2}$& $A_{P}\frac{1}{\sqrt{2}}$ & -12430.53 \\ 
& & $A_{2,2}$&$-A_{P}\frac{1}{2}$ & 4809.16 \\
\cline{2-5}
&$\Omega=0^{-}$ &$A_{5,6}$&$A_{P}\frac{1}{\sqrt{2}}$ & -12462.81 \\
&  &$A_{6,6}$&$-A_{P}\frac{1}{2}$  & 4809.16 \\
\cline{2-5}
 &$\Omega=1$  & $A_{9,10}$ &$-A_{P}\frac{1}{2}$ & 8803.81\\
 & & $A_{9,11}$ & $A_{P}\frac{1}{2}$ & -8812.54\\
&  & $A_{10,11}$&$A_{P}\frac{1}{2}$ & -4812.24 \\
\cline{2-5}
 & $\Omega=2$ & $A_{16,16}$&$A_{P}\frac{1}{2}$  & -4809.16 \\  \hline
\end{tabular}
\caption{Extrapolation of direct spin-orbit couplings $A_{i,j}$ between states correlated to $P+S$ with the functional form $\alpha+\beta_{6}/R^{6}$. $A_{P}=\frac{2}{3}\Delta E_{P}=158.39$ cm$^{-1}$ with $\Delta E_{P}$ being the fine structure splitting of Rb($5p$).}
\label{table:directSOC_P+S}
\end{table}

\begin{table}[]
\centering
\begin{tabular}{c c c c c  } 
\hline
limit& $\Omega$ & $A_{i,j}$ & $\alpha$ & $\beta_{6}$ (a.u) \\ \hline
$S+D$& $\Omega=0^{+}$ & $A_{3,4}$ & $-A_{D}\sqrt{\frac{3}{2}}$ &2139.57 \\
&                & $A_{4,4}$ & $-A_{D}\frac{1}{2}$&  878.59 \\
   \cline{2-5}
   & $\Omega=0^{-}$ & $A_{7,8}$ & $A_{D}\sqrt{\frac{3}{2}}$ &-2151.52\\
   &         & $A_{8,8}$ & $-A_{D}\frac{1}{2}$ & 878.59 \\
   \cline{2-5}
   & $\Omega=1$       & $A_{12,13}$ & $A_{D}\frac{\sqrt{3}}{2}$&-1513.28\\
   &         & $A_{12,14}$ &$A_{D}\frac{\sqrt{3}}{2}$  &-1521.35\\
   &         & $A_{12,15}$ &$0$  &$0$\\
   &         & $A_{13,14}$& $-A_{D}\frac{1}{2}$ & 882.45 \\
   &         & $A_{13,15}$& $-A_{D}\frac{1}{\sqrt{2}}$ & 1254.05 \\
   &         & $A_{14,15}$& $A_{D}\frac{1}{\sqrt{2}}$ & -1254.80 \\
   &         & $A_{15,15}$& $-A_{D}$ & 1787.54\\
   \cline{2-5}
   & $\Omega=2$       & $A_{17,17}$& $A_{D}\frac{1}{2}$ & -878.59\\
   &         & $A_{17,18}$&$A_{D}\frac{1}{\sqrt{2}}$ & -1254.82\\
   &         & $A_{17,19}$ & $A_{D}\frac{1}{\sqrt{2}}$& -1254.80\\
   &         & $A_{18,19}$&$-A_{D}$ & 1787.57\\
   \cline{2-5}
   & $\Omega=3$       & $A_{20,20}$& $A_{D}$  & -1787.54\\
 \hline
\end{tabular}
\caption{Same as Table \ref{table:directSOC_P+S} for the $S+D$ asymptote. $A_{D}=\frac{2}{5}\Delta E_{D}=112.136$ cm$^{-1}$, $\Delta E_{D}$ being being the fine structure splitting of Sr$^+$($4d$)}
\label{table:directSOC_S+D}
\end{table}

\begin{table}[t!]
\centering
\begin{tabular}{c c c c } 
\hline
 $\Omega$ & $A_{i,j}$ & $ \alpha$ & $\beta_{3}$ (a.u)  \\
\hline
 $\Omega=0^{+}$ & $A_{1,4}$& 0.0 & 0.04\\
                & $A_{2,3}$& 0.0  & 0.02 \\ 
                & $A_{2,4}$& 0.0 & 0.0083\\
                 
   \hline
    $\Omega=0^{-}$ & $A_{5,8}$& 0.0 & -0.12\\
                &$A_{6,7}$& 0.0 & 0.11 \\
                & $A_{6,8}$& 0.0 & 0.0083\\
                
   \hline
    $\Omega=1$  & $A_{9,13}$& 0.0 & -0.028\\
                & $A_{10,12}$& 0.0 & -0.012\\
                & $A_{9,14}$& 0.0 & -0.084\\
                & $A_{11,12}$& 0.0 & 0.080\\
                & $A_{9,15}$& 0.0 & 0.0\\
                & $A_{10,14}$& 0.0 & -0.048\\
                & $A_{11,13}$& 0.0 & -0.048 \\
                & $A_{10,15}$& 0.0 & -0.028\\
                & $A_{11,15}$& 0.0 & 0.028\\
   \hline
    $\Omega=2$  & $A_{16,17}$& 0.0 & -0.0083\\
                & $A_{16,18}$& 0.0 & 0.021\\
                & $A_{16,19}$&0.0 & 0.021\\
            
\hline
\end{tabular}
\caption{Extrapolation of indirect spin-orbit couplings $A_{i,j}$ between states correlated to $P+S$ and $S+D$ with the functional form $\alpha+\beta_{3}/R^{3}$.}
\label{table:indirectSOC}
\end{table}

\clearpage

\end{document}